\documentclass[12pt,english,floatfix,nofootinbib,superscriptaddress,aps,prd,preprint]{revtex4}
\usepackage[utf8]{inputenc}
\usepackage{float}
\usepackage{array}
\usepackage{lipsum}
\usepackage{graphicx}
\usepackage{amsmath,amsthm,amsfonts,amssymb}
\usepackage{graphicx}
\usepackage[english]{babel}
\usepackage{color}
\usepackage{tensor}
\usepackage{esint}
\usepackage[dvips]{epsfig}
\usepackage[dvips]{graphicx}
\usepackage{float}
\usepackage{units}
\usepackage{textcomp}
\usepackage{mathrsfs}
\usepackage{amsmath}
\usepackage[makeroom]{cancel}
\usepackage{amssymb}
\usepackage{amsbsy}
\usepackage{amsfonts}
\usepackage{amssymb,mathrsfs,xcolor}
\usepackage{esint}
\usepackage{array}
\usepackage{graphicx}

\usepackage{hyperref}
\hypersetup{
    colorlinks,
    citecolor=blue,
    filecolor=green,
    linkcolor=purple,
    urlcolor=red,
}

\usepackage{slashed}

\newcommand{\ie}{\begin{equation}}
\newcommand{\fe}{\end{equation}}
\newcommand{\se}{\begin{eqnarray}}
\newcommand{\ff}{\end{eqnarray}}

\usepackage{hyperref}
\hypersetup{colorlinks,breaklinks,
			citecolor=[rgb]{0,0.0,1.0},
            urlcolor=[rgb]{0.0,0.0,1.0},
            linkcolor=[rgb]{0,0.5,0.9}}

\begin{document}

\title{Comment on ``A comment on metric vs metric-affine gravity''}

\author{Gonzalo J. Olmo}
\email{gonzalo.olmo@uv.es}
\affiliation{Departamento de Física Teórica and IFIC,
Centro Mixto Universitat de València--CSIC. Universitat
de València, Burjassot-46100, València, Spain}

\author{P. J. Porfírio}
\email{pporfirio@fisica.ufpb.br}
\affiliation{Departamento de Física, Universidade Federal da Paraíba, Caixa Postal 5008, 58051-970, João Pessoa, Paraíba,  Brazil}




\date{\today}

\begin{abstract}

It has been recently claimed in \cite{Lind} that an action of the Einstein-Palatini form plus a torsionless Pontryagin term (multiplied by a constant) represents a counterexample to the conclusions of  \cite{Exirifard:2007da}, namely, that Lovelock gravity is the only case in which  the metric and metric-affine formulations of gravity are equivalent. However, given that the Pontryagin term (multiplied by a constant) can be written as a total D-divergence, it is a textbook matter to realise that the addition of such (or any other) D-divergence only affects at the boundary, leaving invariant the field equations and its solutions, which are those of GR \`{a} la Palatini. We thus conclude that the example provided in  \cite{Lind} is not a valid counterexample of \cite{Exirifard:2007da}.
\end{abstract}

\maketitle


The torsionless metric-affine action in $D=2n$ dimensions considered in \cite{Lind}, 
 \begin{equation}
     S=\int d^{D}x \left(\sqrt{-g}g^{ab}R_{ab}(\Gamma)+\frac{1}{n}\theta \epsilon^{a_{1}a_{2}...a_{D}}R^{i_{1}}_{\,\,\,i_{2}a_{1}a_{2}}(\Gamma)R^{i_{2}}_{\,\,\,i_{3}a_{3}a_{4}}(\Gamma)...R^{i_{n}}_{\,\,\,i_{1}a_{D-1}a_{D}}(\Gamma)\right) \ ,
     \label{ac}
 \end{equation}
 can be written when $\theta$ is constant as 
  \begin{equation}
     S=\int d^{D}x \left(\sqrt{-g}g^{ab}R_{ab}(\Gamma)+\theta\partial_a K^a\right) \ ,
     \label{ad}
 \end{equation}
with $K^a$ as in Eq.(2) of \cite{Lind}. The addition of this fancy total derivative (or of any other) does not have any effect on the field equations \cite{Landau}. Since the scalar $g^{ab}R_{ab}(\Gamma)$ is just the linear in curvature term of Lovelock gravity, the action (\ref{ac}) can still be regarded as included in the equivalence class of Lovelock theories, up to the addition of irrelevant total divergences.  Thus, the example presented in \cite{Lind} cannot be regarded as a counterexample of the conclusions of \cite{Exirifard:2007da}. \\

If the parameter $\theta$ in  (\ref{ac}) is promoted to the status of field,  one obtains that the connection field equations are given by
\begin{equation}
-\nabla_{c}\left(\sqrt{-g}g^{i_{2}b}\right)+\nabla_{d}\left(\sqrt{-g}g^{i_{2}d}\right)\delta^{b}_{c}-2\left(\nabla_{d}\theta\right)\epsilon^{dba_{3}...a_{D-1}a_{D}}R^{i_{2}}_{\,\,\,i_{3}a_{3}a_{4}}(\Gamma)...R^{i_{n}}_{\,\,\,ca_{D-1}a_{D}}(\Gamma)=0 \ ,
\label{fi}
\end{equation}
which naturally boil down to those of GR when $\theta$ is a constant. In the non-constant case,  taking $b=c$ in Eq.(\ref{fi}), one finds $\nabla_{d}\left(\sqrt{-g}g^{i_{2}d}\right)=0$. Plugging this result back into Eq.(\ref{fi}), one arrives at
\begin{equation}
\nabla_{c}\left(\sqrt{-g}g^{i_{2}b}\right)+2\left(\nabla_{d}\theta\right)\epsilon^{dba_{3}...a_{D-1}a_{D}}R^{i_{2}}_{\,\,\,i_{3}a_{3}a_{4}}(\Gamma)...R^{i_{n}}_{\,\,\,ca_{D-1}a_{D}}(\Gamma)=0.
\label{fi1}
\end{equation}
In particular, when $D=4$, Eq.(\ref{fi}) boils down to
\begin{equation}
    \nabla_{c}\left(\sqrt{-g}g^{ab}\right)=-2\left(\nabla_{d}\theta\right)\epsilon^{dba_{3}a_{4}}R^{a}_{\,\,\,ca_{3}a_{4}}(\Gamma),
\end{equation}  
which is similar to the connection equation obtained either in \cite{Sul} or in \cite{Bo} for the torsionless case and by disregarding the homothetic curvature terms. The resulting equations are obviously different from those found in the metric formulation and lead to different phenomenology \cite{Bo2,Bombacigno:2022naf}.  Thus, when $\theta\neq$ constant, the conclusions of \cite{Exirifard:2007da} for the action (\ref{ac}) hold as well. 

Before concluding, we note that \cite{Exirifard:2007da} neither provided an algorithm to find all the solutions for the connection in the metric-affine formulation of Lovelock theories nor ruled out the possible existence of solutions different from the Levi-Civita one \cite{Janssen:2019doc}. 

{\bf Acknowledgments.} 
This work is supported by the Spanish Grant PID2020-116567GB-C21 funded by MCIN/AEI/10.13039/501100011033, the project PROMETEO/2020/079 (Generalitat Valenciana), and by the European Union’s Horizon 2020 research and innovation programme under the H2020-MSCA-RISE-2017 Grant No. FunFiCO-777740. PJP would like to thank CAPES for financial support.




\end{document}